\documentclass[conference]{IEEEtran}
\IEEEoverridecommandlockouts

\usepackage{balance}       
\usepackage{graphics}      
\usepackage[T1]{fontenc}   
\usepackage{txfonts}
\usepackage{mathptmx}
\usepackage[pdflang={en-US},pdftex]{hyperref}
\usepackage{color}
\usepackage{booktabs}
\usepackage{textcomp}
\usepackage{booktabs} 
\usepackage{url}
\usepackage{graphicx}
\usepackage{array}
\usepackage[utf8]{inputenc}
\usepackage{csquotes}
\usepackage{flushend}
\usepackage{enumitem}
\usepackage{tabularx}
\usepackage{balance}
\usepackage{siunitx}
\usepackage{csquotes}
\usepackage{multirow}
\usepackage{float}
\usepackage{array}
\usepackage{microtype}        
\usepackage{ccicons}          

\graphicspath{{figures/}}

\renewcommand{\mkbegdispquote}[2]{\itshape}

\newcommand{\system}{Artemis}
\newcommand{\university}{Technical University of Munich}
\def\tweedback{Tweedback}
\def\zulip{Zulip}
\def\rocketchat{Rocket.Chat}
\def\piazza{Piazza}

\def\moodle{Moodle}
\def\tumjudge{TUMjudge}

\newcommand{\gad}{A}
\newcommand{\gbs}{O}
\newcommand{\pgdp}{P}

\newcommand{\pg}[1]{{\noindent \textbf{#1}}}

\newcolumntype{P}[1]{>{\centering\arraybackslash}p{#1}}

\usepackage{tikz}
\usepackage{lipsum}

\newcommand\copyrighttext{%
  \footnotesize \copyright 2021 IEEE.  Personal use of this material is permitted.  Permission from IEEE must be obtained for all other uses, in any current or future media, including reprinting/republishing this material for advertising or promotional purposes, creating new collective works, for resale or redistribution to servers or lists, or reuse of any copyrighted component of this work in other works.}
\newcommand\copyrightnotice{%
\begin{tikzpicture}[remember picture,overlay]
\node[anchor=south,yshift=20pt] at (current page.south) {\fbox{\parbox{\dimexpr\textwidth-\fboxsep-\fboxrule\relax}{\copyrighttext}}};
\end{tikzpicture}%
}

\def\BibTeX{{\rm B\kern-.05em{\sc i\kern-.025em b}\kern-.08em
    T\kern-.1667em\lower.7ex\hbox{E}\kern-.125emX}}
\begin{document}

\title{An Analysis of Programming Course Evaluations Before and After the Introduction of an Autograder}

\author{
    \IEEEauthorblockN{
        Gerhard Hagerer,
        Laura Lahesoo, 
        Miriam Anschütz,
        Stephan Krusche and
        Georg Groh
    }
    \IEEEauthorblockA{
        Department of Informatics,
        Technical University of Munich\\
        Boltzmannstraße 3, 85748 Garching nearby Munich, Germany\\
        Email: \{gerhard.hagerer,laura.lahesoo,m.anschuetz,krusche\}@tum.de,
        grohg@in.tum.de
    }
}

\maketitle
\copyrightnotice





\begin{abstract}


Commonly, introductory programming courses in higher education institutions have hundreds of participating students eager to learn to program. The manual effort for reviewing the submitted source code and for providing feedback can no longer be managed. Manually reviewing the submitted homework can be subjective and unfair, particularly if many tutors are responsible for grading. Different autograders can help in this situation; however, there is a lack of knowledge about how autograders can impact students' overall perception of programming classes and teaching. This is relevant for course organizers and institutions to keep their programming courses attractive while coping with increasing students.

This paper studies the answers to the standardized university evaluation questionnaires of multiple large-scale foundational computer science courses which recently introduced autograding. The differences before and after this intervention are analyzed. By incorporating additional observations, we hypothesize how the autograder might have contributed to the significant changes in the data, such as, improved interactions between tutors and students, improved overall course quality, improved learning success, increased time spent, and reduced difficulty. This qualitative study aims to provide hypotheses for future research to define and conduct quantitative surveys and data analysis. The autograder technology can be validated as a teaching method to improve student satisfaction with programming courses.


\end{abstract}



\section{Introduction}
\label{sec:introduction}


At the \university{},
the number of freshmen computer science students doubled in recent years and reached more than 2500.
Programming is a crucial skill for their academic and professional careers in engineering and natural and social sciences.
Therefore, many instructors apply autograding to programming exercises to provide immediate feedback to students and lower the manual grading effort.

There are different existing autograders such as WebCat \cite{webcat_demo}, JACK \cite{goedicke201710}, Praktomat \cite{Proktomat_Passau}, GraJa \cite{stocker2013evaluation}, and Artemis \cite{krusche2018artemis}. These systems have been thoroughly evaluated with user studies incorporating quantitative analyses of user behavior and questionnaires.
Also, the impact on learning success in terms of students' grades has been analyzed \cite{mci/Otto2017,krusche2018artemis}. 

However, little is known regarding how autograding relates to student satisfaction with the course and its varied teaching aspects. 
Therefore, course evaluations are a standard method for course organizers and lecturers to understand which parts of the course the students liked and which did not. Especially open-ended comments can be insightful since they contain opinions about emerging course aspects that are not being asked in Likert scale questions. Here, a two-pronged analysis approach using text mining in addition to basic statistics ensures no information is lost. In that regard, topic modeling is an established method in qualitative research to derive theories and hypotheses about how certain factors shape the human experience in specific environments, e.g., how an autograding tool can influence the interaction between students and tutors in teaching sessions\footnote{Tutors are students who have passed the course and take on the role of a student lecturer. Tutors hold tutor groups, in which they support students in working on assignments, answer questions, discuss solutions, help with problems, and grade the assignments based on pre-defined grading criteria.}.

We investigate if there are consistent patterns of how the course evaluations changed when the autograder \system{}
was introduced in three different courses over at least two years, i.e., before and after the intervention. In particular, we focus on the expressed satisfaction with different aspects of the courses according to the students' opinions. 
We are interested in the following research questions:



\def \rqlearning{How did students report on their learning experience in course evaluations, and how did it change?}
\def \rqtutors{How did the reported interaction between students and tutors change?}
\def \rqdifficulty{How did the perceived difficulty of the practical programming parts of the courses change?}
\def \rqprobsolve{How does the students' self-reported capability to solve a problem on their own change?}
\def \rqquality{How did the perceived overall course quality change?}

\begin{enumerate}[label=RQ\arabic*, leftmargin=8mm]
\setlength\itemsep{0em}
\item \rqlearning
\item \rqtutors
\item \rqdifficulty
\item \rqquality
\end{enumerate}

We collected university course evaluations consisting of Likert scale questions and free text comments for three courses.
In each evaluation, 100-500 students, see Table \ref{tab:courses}, provided ratings alongside positive and critical comments about the course implementation, software, teachers, structure, and exercises.
We analyze the evaluation differences before and after the \system{} introduction using statistics and topic modeling.



Section \ref{sec:related-work} provides related work regarding text mining on course evaluation comments and how autograders have been evaluated so far. 
Section \ref{sec:autograder} describes the autograder \system{} and its features in more detail.
In Section \ref{sec:study-design}, we present the study design with the course profiles and the evaluation analysis.
Section \ref{sec:results} presents the results and Section \ref{sec:findings} contains the main findings of this paper.
Finally, section \ref{sec:conclusion} concludes the paper.

\section{Related Work} \label{sec:related-work}

This section motivates why we additionally leverage text mining on course evaluations and why they are a meaningful source for evaluating autograders.

\subsection{Text Mining Applications on Course Evaluations}

Evaluation questionnaires at universities often provide numerical Likert scale questions along with open-ended questions to their participants, such that they can express additional thoughts about the course as free-text comments.
Manually processing them can be labor-intensive, leading to a trend to process comments automatically with text mining for reproducible results \cite{groenbergMSc2020, nasim2017sentiment}. 
This can be applied to multiple courses to harmonize the evaluations and to make the results comparable \cite{HUJALA2020103965,gottipati2017conceptual}. 
The approach carries potential for new applications in educational marketing, e.g., comparing courses on a faculty or university level to improve the overall teaching quality and giving insights about the perception towards an institution \cite{SRINIVAS2019974}.

Relevant text mining techniques for course evaluation analysis are topic modeling and sentiment analysis. These are used in automated tools such as Palaute \cite{10.1145/3428029.3428565,groenbergMSc2020} and SMF \cite{conf/fieitinGS15}, visualizing the underlying topic distribution with non-negative matrix factorization (NMF) or latent Dirichlet allocation (LDA) and students' sentiment \cite{8759085, gottipati2017conceptual}. Even though there is a plethora of more advanced state-of-the-art text mining methods, we do not consider any which have not been previously applied on course evaluations. 
Instead, there are promising method evaluations of the previously mentioned techniques \cite{doi:10.1080/02602938.2020.1805409,HUJALA2020103965, SRINIVAS2019974}, for instance, to depict topic trends of a course over several years \cite{8756781}. This motivates our choice for NMF-based topic modeling applied on course evaluations of multiple courses and years to measure the impact of a new autograding tool. We relate this type of topic distributions in free-text comments with Likert scale answers, as this technique showed statistically significant correlations between both \cite{HUJALA2020103965}. This has the potential to compensate for missing data points and biased open- or close-ended questions. However, the relation between both answer types has not been demonstrated qualitatively, and there is a lack of related applied research regarding the impact of new teaching concepts or tools.

\subsection{Autograder Evaluations}

In this section, we introduce autograding tools and compare how these have been evaluated so far. Evaluation is important as autograders have the potential to improve scaling and fairness, whereas concerns exist regarding the potential lack of personalized feedback and thus decreased motivation.

\subsubsection{Autograders} JACK \cite{goedicke201710} is a web-based autograding tool introduced at the university of Duisburg-Essen. The Praktomat \cite{Proktomat_Passau} autograding tool, used by the KIT and the University of Passau, can grade submissions to multiple languages such as Java, C++ or Haskell. The Web-CAT \cite{webcat_demo} autograder is a very customizable and extensible autograder that is used as a basis for the GraJa \cite{stocker2013evaluation} autograder, for example. Vocareum classroom \cite{vocareum} is a commercial software for software development labs that also includes autograding functionality. Keuning \cite{keuning2018systematic} provides an overview of automated feedback systems, of which a minority are evaluated with questionnaires and a sufficiently large n. Autograding is not mentioned as functionality.

\subsubsection{Perceived Usability}

Given that web-based autograders store the usage data, the according usage patterns can be analyzed. Previous work evaluated questions, such as, if students liked the appearance and how easy the tool was to use \cite{Paredes2017}, how many submissions were graded and which technical issues the students encountered \cite{stocker2013evaluation}, and what the students liked best and what they liked worst in immediate in-class feedback \cite{asee_peer_33235}. While these studies show that autograders can be used successfully by many students in practice to learn programming, it is not clear if and how they contributed to the overall course quality from the students' perspective.




\subsubsection{Learning Experience}

A relevant autograder evaluation aims at testing a gamification concept to improve student engagement in learning programming with Web-CAT \cite{goldman2019using}. The study uses feedback questionnaires about the learning experience, but it does not explain the role or impact of the autograder.

To evaluate the JACK autograder, the overall course evaluation questionnaire \cite{goedicke2008computer} was analyzed, where students referred to the autograder as a positive feature. As only one year of one course is evaluated, it is not clear if the autograder would be seen as beneficial for other courses as well, and it is not compared with when not using an autograder. In a follow-up study, the exam grades were compared to the grades of the previous year that did not use JACK \cite{mci/Otto2017}. They report an improvement in the average grade of students using JACK. Other course aspects, e.g., teaching quality, are not considered.

\section{Autograder} \label{sec:autograder}

\system{} \cite{krusche2018artemis} is a learning management system with individual feedback \cite{krusche2017experiences} that supports interactive learning \cite{krusche2017interactive,krusche2020modeling} and is scalable to large courses \cite{krusche2019mooc}. 
It is open source\footnote{\url{https://github.com/ls1intum/Artemis}}
and used by multiple universities and courses.
It includes autograding functionality to provide feedback to students regarding programming exercises in interactive instructions which change their status and color based on the progress of students.
Completed tasks and correctly implemented model elements are marked in green, incomplete and not yet implemented ones are marked in red.
This helps students to identify which parts of the exercise they have already solved correctly and improves the understanding of the source code on the model level.
When they submit their current solution, the interactive instructions dynamically update.

The programming exercise workflow is as follows:
An instructor sets up a version control repository containing the exercise code handed out to students and test cases to verify students' submissions (\textit{template repository}). 
It includes a small sample project with predefined classes and dependencies to libraries.
The instructor stores the tests for autograding in a separate \textit{test repository}, inaccessible to students.
A combination of behavioral (black-box), structural (white-box) tests and static code analysis allows to check for functionality, implementation details, and code quality of the submission.

After setting up the template, test, and solution repositories, the instructor configures the build plan on the continuous integration server which compiles and tests the exercise code using the previously defined test cases and the static code analysis configuration (\textit{template build plan}).
The build plan includes tasks to pull the source code from the template and test repository whenever changes occur and to combine them so that the tests can be executed in the second step.
A final task, which is executed when compilation or test execution fails, notifies \system{} about the new result.

A student starts an exercise with a single click, triggering the setup process: 
\system{} creates a personal copy of the template and the \textit{student repository} and grants access only to this student. 
It also creates a personal copy of the template build plan and the \textit{student build plan} and configures it to be triggered when the particular student uploads changes to this repository. 
The student can not access the build plan to hide its complexity.
Personalized means that each student gets one repository and one build plan.
When 2,000 students participate in an exercise, \system{} creates 2,000 student repositories and 2,000 student build plans.
Students only have access to their personal repository without access other student repositories.

\begin{table}
\centering
\caption{Courses (\system{} in bold) and number of answers in the data . 
}
\begin{tabular}{lllll}
    \toprule
    \multirow{2}{*}{Course} &  \multirow{2}{*}{\#Answers} & Homework & \multirow{2}{*}{Communication} \\ &&submission&\\ 
    \midrule
    \gad.2019 & 299 & \moodle{}, \tumjudge{} &  \\ 
    \textbf{\gad.2020} & \textbf{269} & \textbf{\system{}} & \textbf{\zulip{}, \tweedback{}} \\ 
    \gbs.2016 & 138 & Git &  \\ 
    \gbs.2017 & 116 & Git &  \\ 
    \gbs.2018 & 182 & \moodle{} & \moodle{} \\ 
    \textbf{\gbs.2019} & \textbf{169} & \textbf{\system{}} & \textbf{\moodle{}} \\ 
    \pgdp.2018 & 519 & \moodle{} & \piazza{} \\ 
    \textbf{\pgdp.2019} & \textbf{553} & \textbf{\system{}} & \textbf{\rocketchat{}} \\ 
    \bottomrule
\end{tabular}
\label{tab:courses}
\end{table}

After the setup, \system{} allows the student to work in a local IDE or in the online editor. 
When the student submits a new solution, the build plan compiles the code and executes the tests defined by the instructor in a docker container.
It uploads the results to \system{}, so that the students can immediately review the feedback and iteratively improve the solution.
In case of an incorrect solution, the feedback shows a message for each failed test.
The student can reattempt to solve the exercise and submit new solutions.
The instructor can review results, gain insights, and react to errors and problems.

\system{} includes a web editor allowing inexperienced students to participate in exercises without dealing with complex version control and integrated development environments. 
It supports the manual review of submissions after the due date.
Tutors can see the automatic feedback trough tests and static code analysis and enhance it with manual feedback.
This makes it possible to review aspects difficult to assess automatically, e.g., the internal structure and specific code quality.

\system{} also features autograding functionality for text exercises \cite{bernius2021machine} and modeling exercises \cite{krusche2022semiautomatic} using a semi-automatic approach based on supervised machine learning.
During the manual assessment, \system{} learns which model elements or text segments are correct or incorrect and applies this information including the qualititative feedback of the manual assessment to similiar model elements or text segments in other students' submissions.
This approach allows multiple correct solutions by students and therfore does not limit their creativity.
The knowledge increases over time and can be reused in subsequent years when exercises are reused.


\section{Study Design} \label{sec:study-design}

At the \university{}, students can evaluate the courses they have taken through anonymous feedback. Questionnaires are pre-defined and distributed every semester for every course by the student council. They contain open-ended text fields where students describe in own words aspects of the course they appreciated and which could be improved. In other questions students rate different aspects of the course on a Likert scale, e.g., lecture materials or homework exercises.

\begin{figure}[t]
\centering
\includegraphics[width=.48\textwidth]{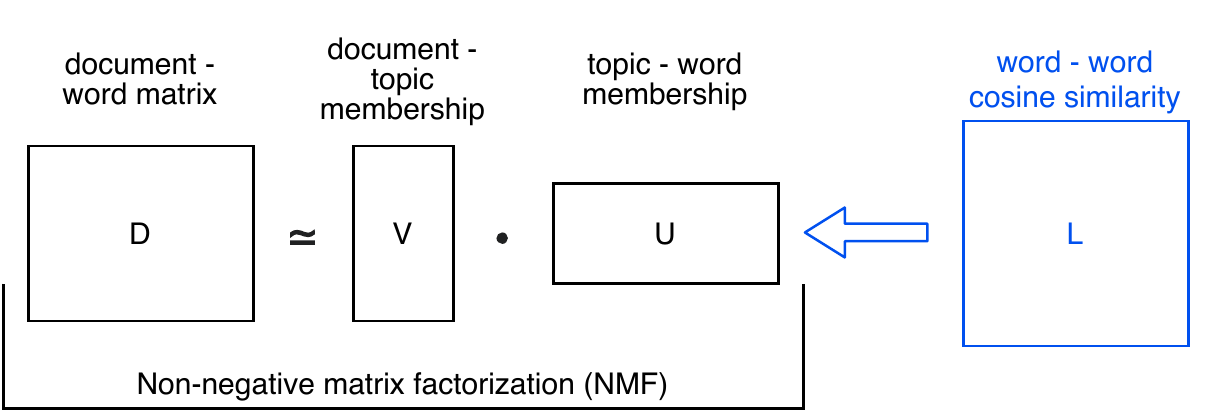}
\caption{Black: Topic modeling based on non-negative matrix factorization (NMF). Blue: Transfer learning using word-word cosine similarities of fastText embeddings. The latter ensures that the NMF and fastText word embeddings correspond to each other. This is referred to as knowledge-guided NMF.}
\label{fig:kgnmf_nmf_vis}
\end{figure}

\newcommand{\multicell}[2][c]{%
  \begin{tabular}[#1]{@{}c@{}}#2\end{tabular}}

{
\newcounter{enrowcount}
\setcounter{enrowcount}{0}
\def\rowcountautorefname{Topic}
\def\rowcountrefname{Topic}
\def\rowlabel{}
\newenvironment{clustering}{}{}

\begin{table*}[t]
\centering
\caption{Top words for each topic translated to English. 
Each comma-separated term represents a single word in the original results. 
The words are sorted in descending order based on the topic-word membership score. 
Anonymized top words are denoted by <>. 
}
\begin{clustering}
\begin{tabular}{>{\rowlabel\space} l p{13cm}}
    \toprule
    \textbf{Topic label} & \textbf{Top words}
    \gdef\rowlabel{\refstepcounter{enrowcount}\label{cluster:\cluster}}
    \gdef\cluster{orga} \\ 
    \midrule
    Course implementation (IMP) & lecture, tutor exercise, hold, synchronize, content, find, work through, lesson, super, book
    \gdef\cluster{samples} \\ 
    Sample solutions (SOL) & homework, sample solution, publish, solution, homework, build upon, helpful, on top of each other, exercise task
    \gdef\cluster{tutoring} \\ 
    Tutoring (TUT) & materials, explain, tutor, understand, understanding, quick, task, detailed, discuss, help
    \gdef\cluster{homework} \\ 
    Homework (HW) & task, week, practical, difficult, explanation, task description, learn, actually, <course \pgdp>, process
    \gdef\cluster{programming} \\ 
    Learning programming (PRO) & learn, programming, practical, application, actually, <\system>, c, apply, practical course, java
    \gdef\cluster{artemis} \\ 
    Homework platforms (ONL) & <\system>, test, run, interactive, hand-in, interaction, work, platform, quiz, <\moodle>
    \\ 
    \bottomrule
\end{tabular}
\label{tab:topwords}
\end{clustering}
\end{table*}
}

\def\orgaref{\hyperref[cluster:orga]{IMP}}
\def\solref{\hyperref[cluster:samples]{SOL}}
\def\tutref{\hyperref[cluster:tutoring]{TUT}}
\def\hwref{\hyperref[cluster:homework]{HW}}
\def\proref{\hyperref[cluster:programming]{PRO}}
\def\onlref{\hyperref[cluster:artemis]{ONL}}

\subsection{Course Profiles}

In the scope of this study, we collected evaluation results from 8 courses held between 2016 and 2020. These belonged to 3 distinct modules of the study plan of informatics, i.e., some courses were repeating instances of the same module held in different years. In the following, we abbreviate each course denoting the module (\gad, \gbs, \pgdp), followed by the year in which the course was held.

Modules \gad{} and \gbs{} are lecture courses accompanied by tutor exercises and homework assignments. Programming tasks are a significant part of the homework in each course. However, good performance in homework assignments is not a central goal for the students, as homework solutions can only yield a grade bonus to improve the final exam grade. 
In contrast, the module \pgdp{} is a practical programming course. There, the student's grade is determined from individually submitted solutions to the programming assignments. Although \pgdp{} is commonly taken together with a introductory informatics lecture, the modules are graded separately.


\subsection{Evaluation Analysis}

The analysis of the questionnaire responses deals with two types of data: On free text replies, the topic distribution is shown for each course, and on Likert scale answers, the mean response rating is computed. We manually map each topic to a related question and plot both together in Section \nameref{sec:results}.

\subsubsection{Topic Modeling on Comments} \label{sec:topic-modeling}

In all programming course evaluation questionnaires, participants write free-text answers to two questions: a) Which aspects of the course did you appreciate? (positive); b) What should be improved? (negative). With topic modeling \cite{finch2018use}, these comments are clustered into semantically coherent topics. This shows the themes of interest for the students.

We use non-negative matrix factorization (NMF), an established technique for topic modeling \cite{cichocki2009fast}.
NMF takes a document-word matrix $D$ as input, in which each row represents a document, each column a word, and each cell the term frequency-inverse document frequency (TF-IDF). It is a weighted count of word occurrences within a document, such that a word occurring in all documents gets a low score \cite{salton87}. The matrix $D$ is factorized into the document-topic matrix $V$ and the topic-word matrix $U$ -- see Figure \ref{fig:kgnmf_nmf_vis}. The scores in $V$ and $U$ show how much a document or word is related to a topic. As we are working with few short texts, $D$ becomes sparse, leading to semantically incoherent word lists for the extracted topics. Therefore, we include external knowledge about the words, i.e., their similarities according to external corpora, by using knowledge-guided NMF (KG-NMF) \cite{KGNMF_orig}.




\subsubsection{Likert Scale Questions} \label{sec:likert-scale-questions}

In the standardized evaluation, Likert scale questions can have one of three formats: a) Students are asked to state a grade between 1 ``best'' and 5 ``worst''; b) students evaluate their agreement with a given statement ranging from 1 ``fully agree'' to 5 ``completely disagree''; c) students can pick a position on a five-point scale between two extremes, e.g., the speed of a lecture could be 1 ``too fast'' and 5 ``too slow'', with 3 implying ``just right''. 


\def \qtotalgrade{\textit{``Which overall grade would you give this course?''}}
\def \qtutorgrade{\textit{``Which grade would you give your tutor?''}}
\def \qsamplesols{\textit{``All in all, the homework sample solutions are helpful.''}}
\def \qhwdifficulty{\textit{``The difficulty level of the homework exercises is adequate.''}}
\def \qproblemsolve{\textit{``I have learned to solve problems typical for the course's domain.''}}
\def \qonlineoffer{\textit{``The online offering for the course is good.''}}
\def \qtimehomework{\textit{``On average, how many hours per week do you spend on homework assignments?''}}

The Likert scale answers to the following survey questions and statements are discussed in the \nameref{sec:results}:

\begin{itemize}
\setlength\itemsep{0em}
    \item \qtotalgrade{}
    \item \qtutorgrade{}
    \item \qsamplesols{}
    \item \qhwdifficulty{}
    \item \qproblemsolve{}
    \item \qonlineoffer{}
\end{itemize}

As exact phrasing of questions varied across years, synonymous questions are grouped together manually. Some questions are removed from the questionnaires by some lecturers, since the standardized university evaluation can be modified every year. The questionnaire for \gad.2020 especially focused on the challenges of digital study and omitted standard questions.

\subsection{Addressing the Research Questions}

The research questions in the \nameref{sec:introduction} are concerned with the impact of the \system{} autograder on the students' experience of programming courses. We analyze the impact by qualitatively comparing course evaluation statistics before and after introducing the tool. The statistics are a) the averaged Likert scale answers to the questions in subsection \ref{sec:likert-scale-questions}, and b) corresponding topic distributions of comments about positive and negative aspects in the courses -- see Table \ref{tab:topwords}.

Each research question is related to an evaluation question and topic, such that the answers and topic distributions together provide a meaningful answer. For a topic and a question about the same teaching aspect, we also depict if the topic appearance is reflected in the mean question response. If, e.g., the positive comments and the Likert answers increase over the years, we would conclude the regarding course aspect improved. 




\section{Results} \label{sec:results}

The results of the course evaluation questionnaires are shown as diagrams. Each consists a bar and a scatter plot. The scatter plot depicts the average of all numerical answers to the close-ended question of a course year. Values of the same course module over years are connected with lines. The last year of each module is highlighted, as this was the first time when \system{} was used in that course. To this year the p-value of a significance test is annotated\footnotemark. The smaller the number, the less likely it is that both distributions are equal. As in similar studies \cite{maki2000evaluation,lake2001student,hoag2005piloting}, we accept p-values $<0.05$ as statistically significant, which holds for all measured p-values.

The bar plots are obtained by summing up the comment-topic-membership scores over the positive or negative comments of the given topic. 
The highest-scoring top words of each topic are in Table \ref{tab:topwords}. We labeled each relevant topic manually based on top words and comments. The number of topics for NMF is 13, after evaluating the top word coherence manually for different topic counts. From these, 6 relevant topics are chosen. These are mapped to the most related questions and both are shown together in the diagrams and subsections.




\subsection{Course Quality}
\label{sec:course-evaluation}

\begin{figure}[!h]
\vspace{-4mm}
\centering
\includegraphics[width=\columnwidth]{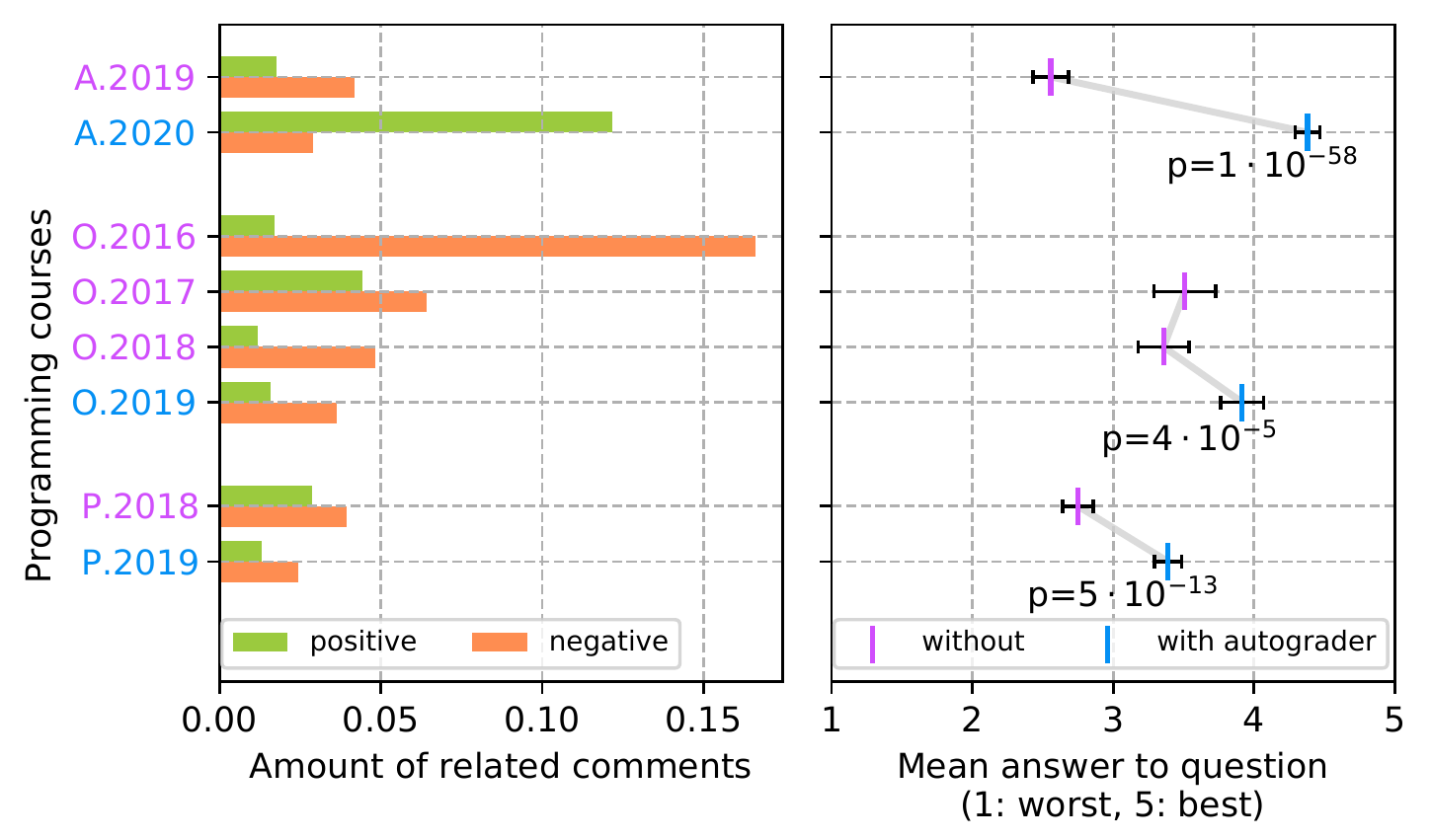}
\vspace{-8mm}
\caption{Left: Number of positive/negative comments on topic \orgaref{}. Right: mean of numerical answers with black confidence intervals\textsuperscript{\ref{ft:conf-interval}}; annotated p-values\textsuperscript{\protect\ref{ft:test}}.}
\label{fig:orga}
\vspace{-3mm}
\end{figure}

\pg{Question:} \qtotalgrade{}

\pg{Topic:} \orgaref{} collects general feedback about the lectures and tutor exercises held during the course. In the topic's comments, students frequently commented about the quality of the course and the themes as these are discussed in lectures and the related tutor exercises. The top words of the cluster highlight the connection between \textit{lectures} and \textit{tutor exercises}. 

\pg{Results:} \autoref{fig:orga} shows that all courses received a higher rating after the introduction of \system{}. 
Simultaneously, either the negative comments about the course implementation receded or positive comments increased.

\subsection{Tutoring}
\label{sec:tutoring}

\begin{figure}[h]
\vspace{-4mm}
\centering
\includegraphics[width=\columnwidth]{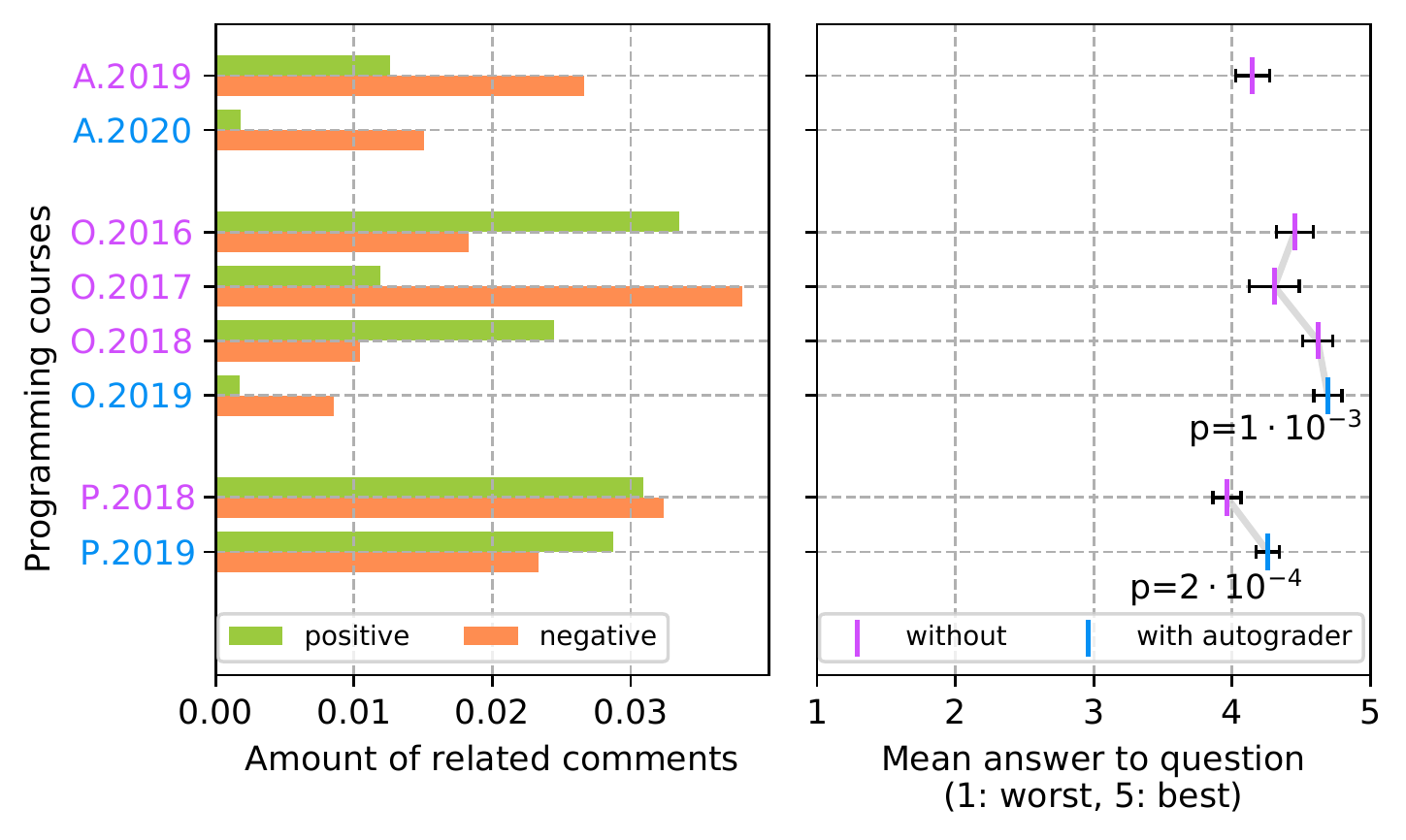}
\vspace{-8mm}
\caption{Left: Number of positive/negative comments on topic \tutref{}. Right: mean of numerical answers with black confidence intervals\textsuperscript{\ref{ft:conf-interval}}; annotated p-values\textsuperscript{\protect\ref{ft:test}}.}
\label{fig:tutoring}
\vspace{-3mm}
\end{figure}

\pg{Question:} \qtutorgrade

\pg{Topic}: \tutref{}'s top words reflect the interactions between the students and tutors, where we find \textit{explain, understand, detailed, discuss}, and \textit{help}.

\pg{Results:} 
The rating for the tutor was higher after the introduction of \system, see the courses \gbs{} and \pgdp{} in \autoref{fig:tutoring}. 
The number of negative comments corresponds inversely with the grade students gave for their tutors. 
This shows that the tutors were perceived increasingly positively after Artemis was used. The course \gad{}.2020 questionnaire omitted this and other questions, such that no comparisons can be made for it.

\subsection{Sample Solutions}
\label{sec:samples}

\begin{figure}[h]
\vspace{-5mm}
\centering
\includegraphics[width=\columnwidth]{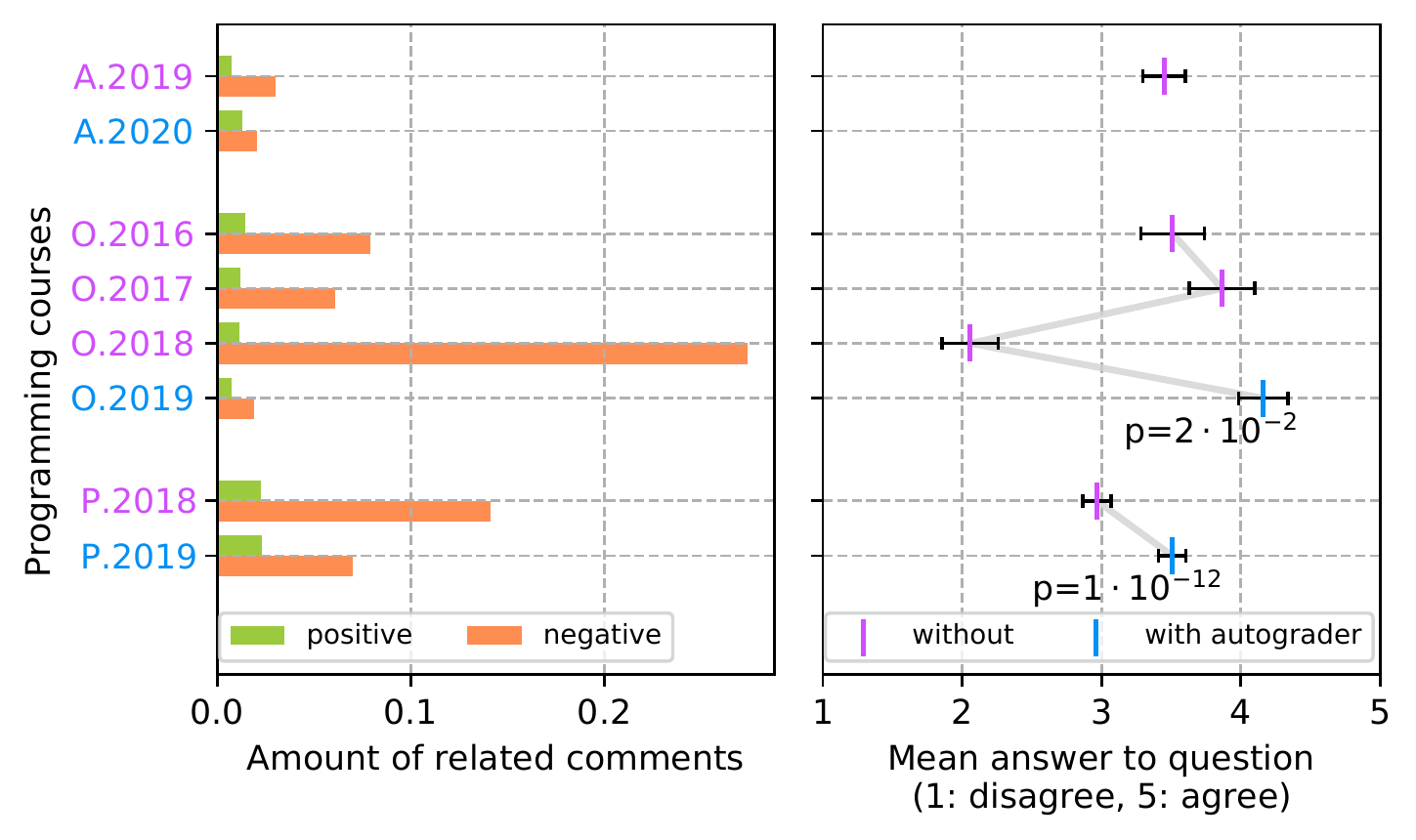}
\vspace{-8mm}
\caption{Left: Number of positive/negative comments on topic \solref{}. Right: mean of numerical answers with black confidence intervals\textsuperscript{\ref{ft:conf-interval}}; annotated p-values\textsuperscript{\protect\ref{ft:test}}.}
\label{fig:samples}
\vspace{-3.5mm}
\end{figure}

\pg{Question:} \qsamplesols{}

\pg{Topic:} \solref{} collects feedback on \textit{homework} exercises with an emphasis on the \textit{publishing} of \textit{sample solutions}, which \textit{build upon} and \textit{on top of each other}. 

\pg{Results:} Course \gbs.2018 received many negative \solref{} comments and a low rating. According to the interviews with tutors in section \ref{sec:tutor-improvement}, 
students had difficulties with homework assignments built upon solutions of previous weeks, which were not published in time and thus hindered them from solving those assignments. Apart from that, sample solutions overall were rated more positively and commented less negatively after the \system{} intervention.


\subsection{\system} \label{sec:system}

\begin{figure}[h]
\vspace{-3mm}
\centering
\includegraphics[width=.99\columnwidth]{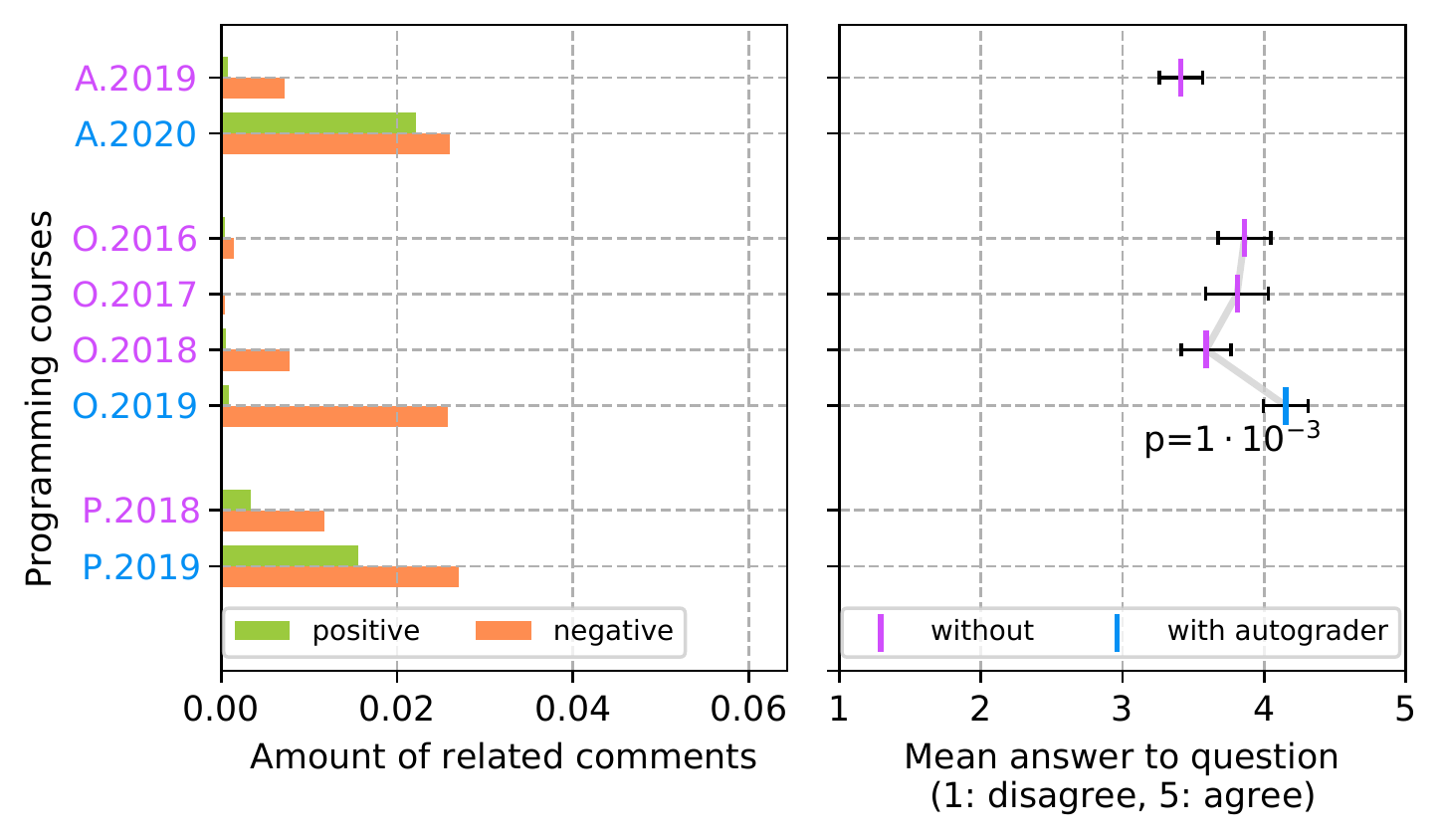}
\vspace{-5mm}
\caption{Left: Number of positive/negative comments on topic \onlref{}. Right: mean of numerical answers with black confidence intervals\textsuperscript{\ref{ft:conf-interval}}; annotated p-values\textsuperscript{\protect\ref{ft:test}}.}
\label{fig:artemis}
\vspace{-3.5mm}
\end{figure}

\pg{Question:} \qonlineoffer

\pg{Topic:} \onlref's top words regard to student interactions with online platforms (\textit{\system, \moodle}) and their functionalities (\textit{work, run, test, hand-in, quiz}). 

\pg{Results:} With the introduction of \system{}, the rating improved in \gbs{}.2019 and the overall number of \onlref{} comments increased. 
In the negative comments for Topic \onlref{}, the students frequently provided bug reports or change requests for the newly introduced \system. As a consequence, the negative comments from the students do not depict strong negative sentiment, but constructive criticism, which can be seen on the Likert scale answers.

\subsection{Homework Exercises}
\label{sec:homework}

\begin{figure}[h]
\centering
\vspace{-2mm}
\includegraphics[width=\columnwidth]{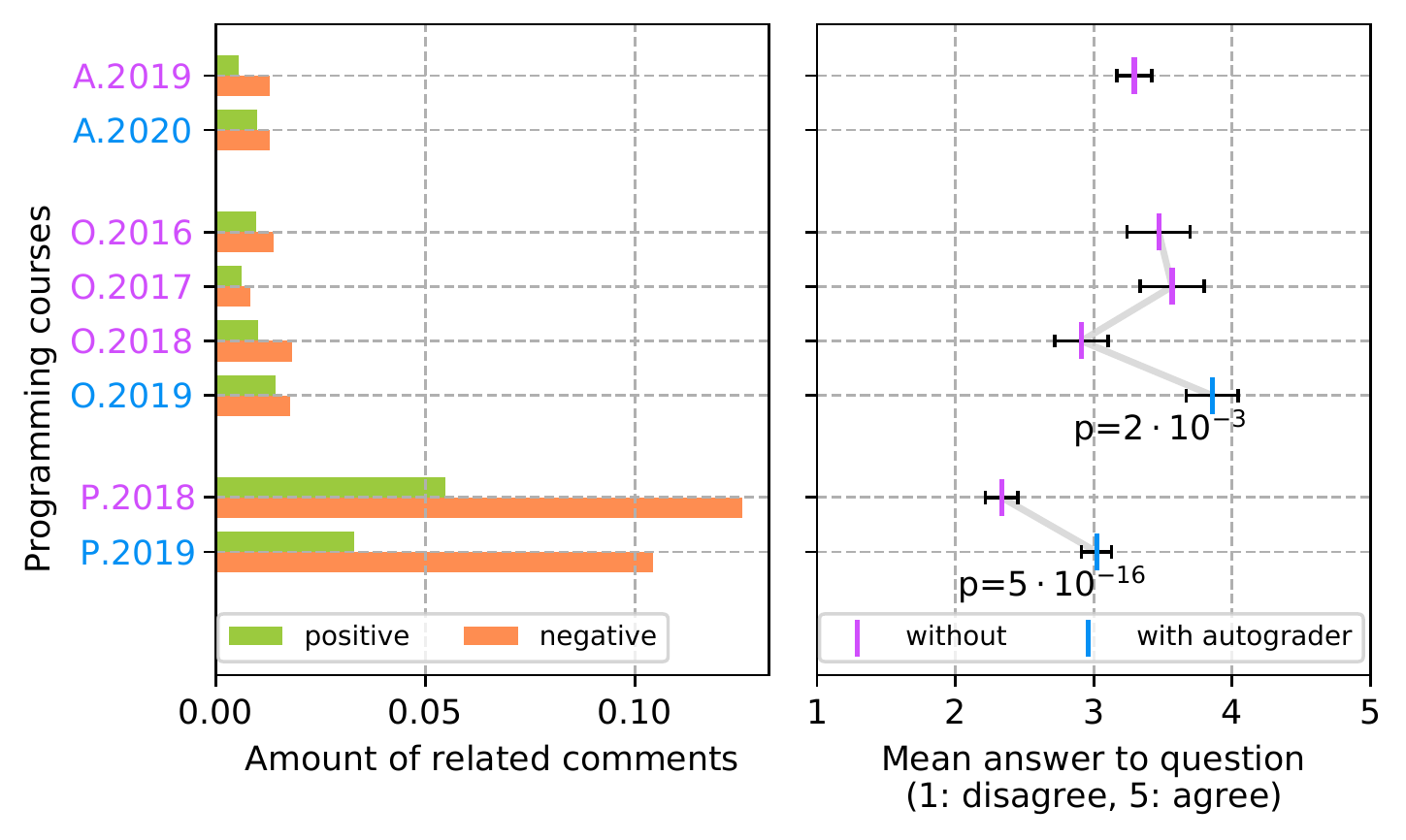}
\vspace{-8mm}
\caption{Left: Number of positive/negative comments on topic \hwref{}. Right: mean of numerical answers with black confidence intervals\textsuperscript{\ref{ft:conf-interval}}; annotated p-values\textsuperscript{\protect\ref{ft:test}}.}
\label{fig:homework}
\vspace{-2mm}	
\end{figure}


\pg{Question:} \qhwdifficulty

\pg{Topic:} \hwref{} is about \textit{weekly} \textit{tasks} meant to help students \textit{learn} course contents \textit{practically}. In \autoref{fig:homework}, the topic appeared primarily in the \pgdp-courses, which are focused on extensive homework based on which the complete grade is calculated. 
In other courses, good homework performance gives a small bonus on the exam grade. 

\pg{Results:} 
Except for the denoted outlier \gbs.2018, homework assignments are perceived as more adequate by each consecutive year in \autoref{fig:homework}. The negative comments of the homework topic are distributed similarly for the course series with sufficient data-points, i.e., \gbs{} and \pgdp{}. However, the correspondence with positive comments is not clear for \pgdp{}.



\subsection{Time Consumption of Homework}
\label{sec:timeconsumption}


\pg{Question:} \qtimehomework

\pg{Results:} As no topics and only incomplete data are available, we only provide the average numbers for the course \gbs{} as follows.
The average time consumption of homework exercises remained constant at 3 hours untill 2017, increased to 6 hours in 2018 due to \autoref{sec:samples}, and decreased to 4 hours in 2019.
This means that in 2019, when \system{} was introduced, the reported time spent on homework was higher than in 2016 and 2017. 
Concurrently, there was an increase in the course satisfaction (\autoref{fig:samples}), the adequateness of the homework tasks (\autoref{fig:homework}), and the learning outcome (\autoref{fig:programming}). Thus, it cannot be excluded that \system{} led to higher motivation and increased time spent on homework.
However, the data is only available for one course series, and further research should investigate if other courses also show increased time for solving homework assignments, especially in the long run.

\footnotetext[5]{\label{ft:conf-interval}These are confidence intervals with a 95\% chance of containing the true population mean answers. We have $n>100$ samples (see Table \ref{tab:courses}) and assume a normal distribution for the means due to the central limit theorem.}

\footnotetext{\label{ft:test}Null hypothesis: Both distributions are equal. Alternative hypothesis: The distribution from the year with autograder is higher than from the years before. A one-sided Wilcoxon rank-sum test \cite{Wilcoxon1992} is applied, since the distributions are not normally distributed, and the years with autograder have higher ratings. For the sample sizes see Table \ref{tab:courses}.}

\subsection{Learning Programming}
\label{sec:programming}

\begin{figure}[h]
\vspace{-0.94em} 
\centering
\includegraphics[width=\columnwidth]{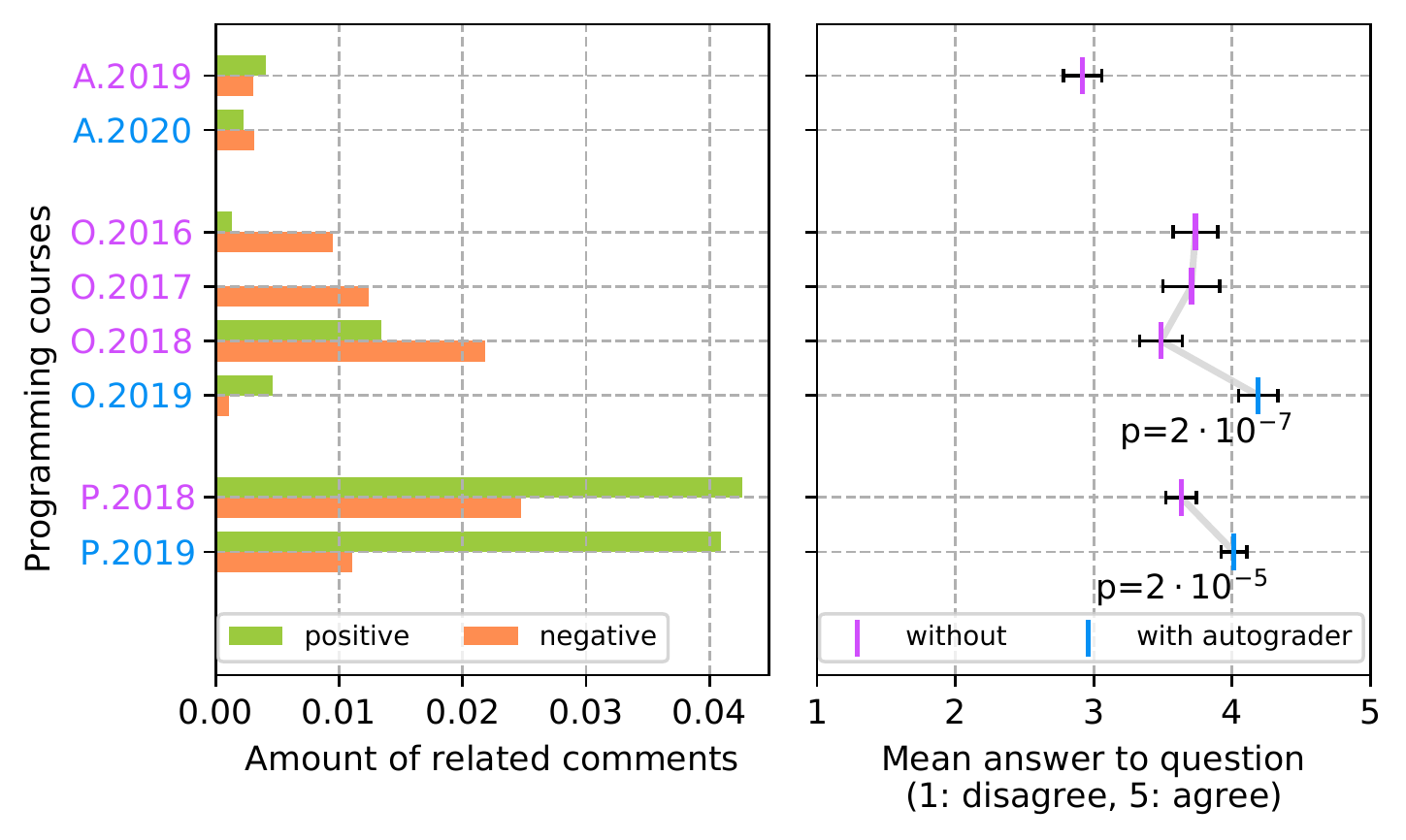}
\vspace{-8mm}
\caption{Left: Number of positive/negative comments on topic \proref{}. Right: mean of numerical answers with black confidence intervals\textsuperscript{\ref{ft:conf-interval}}; annotated p-values\textsuperscript{\protect\ref{ft:test}}.}
\label{fig:programming}
\vspace{-2.7mm}
\end{figure}

\pg{Question:} \qproblemsolve

\pg{Topic:} \proref{} collects feedback on the development of programming skills. Its top words relate to \textit{programming}, programming languages (\textit{C, Java}), \textit{applying practical} skills, and \textit{application} development. \textit{\system} is also among them, indicating a connection with learning programming in students' feedback.

\pg{Results:} The students' perceived programming learning success was rated highest for both courses \pgdp{} and \gbs{} in the year when \system{} was introduced. Similarly and at the same time, the number of negative comments in \autoref{fig:programming} decreased much more than the positive ones. The practical course \pgdp{}, where students are mainly focused on solving programming tasks, shows the highest proportion of comments in \autoref{fig:programming}.

\subsection{Further Observations}
\label{sec:further-observations}

In addition to the results shown above, further data shows positive trends in years when \system{} was introduced. Where comparison data is available, we observe the students' average numerical response improving for any other question related to tutor exercises, homework, or student learning progress. Such questions and prompts include:

\begin{itemize}
\setlength\itemsep{0em}
    \item \textit{The difficulty of tutor exercise content is …}
    \item \textit{I can now provide an overview about the topic of the course.}
    \item \textit{I can now explain important concepts from the course.}
    \item \textit{How well were you able to acquire the required knowledge using the provided materials?}
\end{itemize}

In all these cases, we observe that modules have a higher numeric score for a course with \system{} compared to all other scores within the module for the same question.

\section{Findings} \label{sec:findings}

\subsection{Answers to Research Questions}

The research questions from the
\nameref{sec:introduction}
regarding how the course evaluation surveys changed after the introduction of \system{} can be answered as follows. 


\subsubsection{RQ1 \rqlearning} 

The students' perception of their ability to solve domain-typical programming problems increased in all courses with according data (\ref{sec:programming}), while they spent more time on the exercises (Section \ref{sec:timeconsumption}), potentially because of improved reported learning experiences (Section \ref{sec:further-observations}).


\subsubsection{RQ2 \rqtutors} 

Section \ref{sec:tutoring} shows tutors and their competences are perceived more positively by students. Supporting arguments are given by interviews we conducted with tutors and by the related work -- see the section \ref{sec:tutor-improvement}.

\subsubsection{RQ3 \rqdifficulty} 

From Section \ref{sec:homework}, it can be concluded that the difficulty of the programming exercises overall was seen as more adequate by the course participants than it was before after \system{} was introduced.


\subsubsection{RQ4 \rqquality}

The overall rating of all courses increased and the number of critical comments decreased after the introduction of \system{} as outlined Section \ref{sec:course-evaluation}. Potential reasons include the previously reported finding, i.e., improved interactions between tutors and students, improved overall course quality, and improved learning success.



\subsection{Hypotheses About the Impact of \system}

In the following paragraphs, we connect the above findings with several other factors, including the \system{} autograder. This is supposed to inform hypotheses regarding how autograding can contribute to improving student satisfaction in programming courses.

\subsubsection{Higher Motivation for Homework}

After the introduction of \system, students reported that they spent more time on homework exercises, even though they perceived the exercises to be less difficult compared to students who did not have an autograder. 
It appears that the learning experience is more rewarding when homework is solved the homework. Accordingly, the course satisfaction and learning outcome is increased, too. This can be observed across all courses after the introduction of \system.

In this context, it is worth mentioning that unit tests are provided on \system, which are triggered when uploading the homework exercises. The output of the tests become immediately visible to the student. This motivates the student to upload an improved version of the code before the deadline. So, although it seems the student spends more time solving the homework, the feedback loop keeps motivation high. This is confirmed by the literature \cite{keuning2018systematic} and some of the evaluation comments.




\subsubsection{Better Tutoring Sessions}
\label{sec:tutor-improvement}

In Section \nameref{sec:tutoring} we state \textit{``the rating for the tutors was consistently higher after the introduction of \system''}. 
To explain this observation, we interviewed tutors of the course P from both years about the advantages of \system{} during their teaching sessions. They stated or confirmed the following statement:

\begin{displayquote}
``What is very helpful, however, are the automatic tests for the in-class exercises. This means that students do not have to test their code so extensively themselves (especially if it works). In the past, they often needed help or support with these tests. As a result, I have more time for the people who really need help.''
\end{displayquote}

Similar feedback was reported in another paper:

\begin{displayquote}
``Finally, this methodology allows the professor to make better use of her time. Thanks to the use of the automated tool, her time is mostly spent with the students who have not been able to obtain correct answers; the rest of the students already know their solutions are correct, and they have
moved on to more challenging problems.''~\cite{asee_peer_33235}
\end{displayquote}

With the integration of \system, more advanced students can work more independently and thus progress faster. They need less feedback -- especially at the beginning to get going -- as this can be given by automatic tests. Consequentially, tutors can devote their limited time to help the less advanced students. 
We conclude that the usage of an autograding system can have a positive impact on the teaching quality in tutor sessions. This is a possible explanation why tutors who used an autograder received a more positive evaluation.

\subsubsection{Fairer Homework Corrections}

\system{} assigns homework submissions from students
randomly to tutors for correction. 
The process is anonymized and the students do not know which tutor corrected their submissions. This feature is called double-blind homework correction and might have impacted the evaluations. It was a new practice for all given courses, as previously one tutor was always responsible for the homework correction of one fixed group of students throughout the whole semester. This tutor was also known to them as their responsible supervisor. With the new approach, the overall homework points should depend less on the bias of one single tutor. This can reduce unfair treatment in the homework corrections and impact how students evaluate tutors, practical programming parts, or the overall course implementation.

\subsection{Limitations}

The data of this study is not gathered under controlled conditions. 
This means that in addition to \system, there are other factors that impact the student satisfaction. These factors have different degrees of impact and can change from year to year. In the following, we list factors that can impact the satisfaction of users to a high degree. 
However, it should be kept in mind that all teaching courses we take into account in this study are foundational computer science courses. This means that their content and organization do not significantly change.

The year 2018 of the course \gbs{} was evaluated worse than all other years in several categories, for example, clarity, amount, and speed of content (the last two are not discussed in the results). As mentioned in section \ref{sec:samples}, this was caused by an organizational problem with homework assignments. One open question is how this damaged the overall course impression of the students, or if not, which other factors played a role in the ratings. We address the problem by omitting evaluations from the course \gbs{}, which significantly deviate negatively for each Likert question when significance tests are calculated. Consequently, the mean answers for \gbs{} with and without \system{} are more similar but always statistically significantly different with all p-values $<0.05$.

Even though a large sample size of more than 100 evaluation participants (see \autoref{tab:courses}) should give robustness against single-student bias, sympathetic or otherwise highly regarded professors or tutors might have increased positive valuation from students even for unrelated questions \cite{personalityAndSET}, especially as the professors changed in every course from year to year.


The introduction of \system{} frequently coincides with better numerical ratings from students in the presented lectures. A causal link, however, cannot be established without taking a look at lectures or tutor sessions from the same years without \system{}. This would rule out the possibility that the courses are improving each year due to unrelated reasons.

Finally, courses with \system{} should be investigated for selection bias. Since changes to the technical solution of a course require additional effort, it is possible that only course organizers putting more effort towards the course would consider transitioning to a new LMS. Positive factors that arise from motivated course organizers (e.g., better study materials, quicker answering to student questions) might cause higher scores in student evaluations unrelated to \system{}.

\section{Conclusion} \label{sec:conclusion}

In this paper, significant changes in the evaluations of several foundational computer science course series with numerous participants (> 1000) are depicted after introducing an autograding software. Regarding most research questions, consistent improvements from the perspective of the course participants are observed. These include improved interactions between tutors and students, improved course quality, improved learning success, increased time spent, and reduced difficulty. As possible reasons, we identify helpful automated feedback from unit tests, fairer and more objective grading, reduced correction bias, enhanced course implementation, and more available time for tutors to focus on students’ needs during teaching sessions. 

The analysis is based on a qualitative interpretation of statistics and topic modeling. While the measures provide mostly coherent results, the conclusions need to be confirmed by applying thorough quantitative correlation measures, e.g., as conducted in other studies \cite{HUJALA2020103965}. Furthermore, we aim to include sentiment analysis, as the sentiment dimension differs from the positive vs. improvable categories given in the questionnaires used in this study. Regarding the data, we want to include more data to see if the changes of Artemis are sustainable after more than one year.

\bibliographystyle{IEEEtran}
\bibliography{literature}

\end{document}